\documentclass{article}
\usepackage{arxiv}
\usepackage[utf8]{inputenc}
\usepackage[T1]{fontenc}
\usepackage{hyperref}
\usepackage{url}
\usepackage{booktabs}

\usepackage{microtype}
\usepackage{graphicx}
\usepackage{float}
\usepackage[section]{placeins}
\usepackage[numbers,square,sort&compress]{natbib}
\usepackage{doi}

\title{When AI Should Not Decide: Evidence Sufficiency and Decision Stability in Automated Job Interviews}
\renewcommand{\headeright}{}
\renewcommand{\undertitle}{}
\author{Khalid Adnan Alsayed\\
Ducaltus Ltd, London, UK\\
University of Leicester, Leicester, UK\\
\href{https://orcid.org/0009-0000-6272-6761}
{ORCID: 0009-0000-6272-6761}\\
\texttt{Research@ducaltus.com}\\
\url{https://www.ducaltus.com}}
\date{}
\renewcommand{\shorttitle}{When AI Should Not Decide}
\hypersetup{pdftitle={When AI Should Not Decide: Evidence Sufficiency and Decision Stability in Automated Job Interviews},pdfauthor={Khalid Adnan Alsayed},pdfkeywords={AI assurance, automated interviews, evidence sufficiency, decision stability, abstention}}
\begin{document}
\raggedbottom
\setlength{\intextsep}{8pt plus 2pt minus 2pt}
\maketitle

\begin{abstract}
Artificial intelligence is increasingly used to assess employment candidates, yet consistent interview scoring does not necessarily establish whether an automated recommendation is adequately supported by the available evidence. This study introduces the Evidence Sufficiency and Decision Stability Assessment (ESDSA), a framework for examining when LLM-based competency evaluations should produce actionable recommendations or defer further assessment. ESDSA combines criterion-level evidence classification, deterministic quotation traceability and decision-stability checks across linguistic variants assessed as substantively equivalent. The framework was evaluated using 540 synthetic interview responses from 60 evaluated scenario families across Software Engineering, Customer Support and Operations, comprising developmental, validation and held-out test evaluations. Four approaches were compared: scoring-only, evidence-aware, traceability-based and integrated ESDSA v2 assessment. In held-out testing (216 responses), integrated ESDSA achieved 85.19\% actionable coverage and 3.80\% disagreement among actionable recommendations against AI-assisted provisional references, compared with 92.59\% coverage and 4.50\% disagreement for evidence-aware evaluation. The scoring-only baseline achieved lower disagreement (1.95\%) but substantially lower coverage (71.30\%). The integrated stability gate additionally deferred 11 previously actionable decisions, ten of which had agreed with provisional references. These findings reveal a measurable trade-off between evidential constraints, decision consistency and abstention, rather than demonstrating overall superiority of the integrated framework. ESDSA contributes a reproducible methodology for assessing the procedural justification of automated interview recommendations, while recognising that synthetic responses and AI-assisted references do not establish real-world recruitment validity.
\end{abstract}

\section{Introduction}
Artificial intelligence is increasingly being investigated for employment assessment, including the evaluation of open-ended interview responses. Large language models (LLMs) offer opportunities to apply structured competency criteria at scale, yet their ability to generate consistent scores does not establish whether individual recommendations are supported by sufficient evidence. Recent investigations of LLM-based interview scoring have examined assessment validity, model selection and prompting strategies \citep{stockdale2026}. A further question remains: when does the information provided by a candidate justify an actionable recommendation, and when should an automated evaluator defer?

This distinction reflects a longstanding concern in assessment theory. Validity depends not merely on measurement consistency but on the evidential justification of interpretations and decisions \citep{messick1995}. Research on algorithmic hiring likewise identifies discrepancies between assessment claims and the validation evidence supporting them \citep{raghavan2020}. In competency-based interviews, an incomplete answer may fail to demonstrate a particular ability without establishing that the candidate lacks it. Conversely, an answer may explicitly describe materially deficient practice. Treating missing evidence and negative evidence as interchangeable risks converting uncertainty about an interview response into an unjustified judgment about the individual.

Decision stability presents a related challenge. Candidates may express substantively equivalent competency evidence through different linguistic styles, levels of detail or conversational structures. Behavioural testing has demonstrated that language models can exhibit unexpected sensitivity to controlled input variation \citep{ribeiro2020}, while research on LLM-based evaluation identifies instability arising from assessment conditions \citep{anghel2025}. However, communication can itself constitute a legitimate job requirement. A defensible stability assessment must therefore distinguish linguistic differences that preserve the assessed evidence from changes that materially affect its meaning.

These concerns also have institutional significance. The European Union's Artificial Intelligence Act identifies recruitment-related candidate evaluation among the high-risk applications described in Annex III \citep{eu2024}. The NIST AI Risk Management Framework similarly emphasises the governance, measurement and management of AI-related risks \citep{nist2023}. Although neither establishes a specific evidential threshold for individual interview recommendations, both provide context for examining how automated assessments are justified and controlled.

This study introduces the Evidence Sufficiency and Decision Stability Assessment (ESDSA), a reproducible experimental methodology combining criterion-level evidence classification, deterministic quotation traceability and decision-stability checks across linguistic variants assessed as substantively equivalent. The framework distinguishes positive, negative and insufficient evidence and applies predefined rules to determine whether an automated recommendation should ADVANCE, DO\_NOT\_ADVANCE or DEFER. These outcomes represent experimental competency assessments, not actual hiring decisions.

ESDSA was evaluated using synthetic interview responses across Software Engineering, Customer Support and Operations. Following developmental refinement, the frozen ESDSA v2 procedure was compared with scoring-only, evidence-aware and traceability-based approaches in VALIDATION and held-out TEST evaluations. In TEST, integrated ESDSA achieved 85.19\% actionable coverage and 3.80\% disagreement against AI-assisted provisional references, compared with 92.59\% coverage and 4.50\% disagreement for evidence-aware evaluation. The scoring-only baseline achieved lower disagreement (1.95\%) at lower coverage (71.30\%). Moreover, ten of the eleven additional decisions deferred by the stability gate had previously agreed with provisional references.

These findings do not establish superior recruitment accuracy or decision validity. Instead, they expose the trade-offs associated with imposing evidential and stability requirements on automated recommendations. ESDSA's contribution lies in specifying and evaluating those requirements within an interview-focused decision protocol, making the relationship between evidence sufficiency, reference disagreement, decision coverage and abstention explicit.

\section{Related Work and Research Gap}
\subsection{Automated Interview Assessment and Construct Validity}
Research on automated employment assessment increasingly examines whether AI-generated ratings possess the measurement properties required for personnel selection. \citet{stockdale2026} investigated how model selection, prompting and evaluation design influence LLM-based interview scoring, while \citet{hickman2025} examined the susceptibility of automated interview ratings to applicant impression management. These studies contribute to understanding assessment reliability, but consistency with human ratings does not independently establish that a particular recommendation is justified by the candidate's response.

This distinction has established foundations in personnel psychology. \citet{messick1995} conceptualised validity in terms of the evidential justification of assessment interpretations and their uses. Research on structured interviews and personnel selection further demonstrates the importance of assessment design, predictive validity and the combination of information in selection decisions \citep{campion1997,schmidt1998,kuncel2013,levashina2014}. However, standardised scoring cannot by itself establish whether an individual response provides adequate evidence of the assessed competency, particularly where the relationship between the measured construct and its operational representation remains uncertain \citep{jacobs2021}.

Similar concerns arise in algorithmic recruitment. \citet{raghavan2020} questioned the validation practices and evidential claims associated with algorithmic hiring systems, while \citet{aizenberg2025} examined assumptions underlying automated representations of candidate attributes. For competency-based interviews, an additional distinction is necessary: failing to demonstrate competence is not equivalent to demonstrating deficient practice. Evidence sufficiency therefore concerns what can reasonably be inferred from the available response, rather than the candidate's underlying professional ability.

\subsection{Reliability, Linguistic Variation and Evaluator Instability}
Language-model reliability cannot be established through aggregate scoring agreement alone. \citet{ribeiro2020} introduced CheckList, demonstrating how targeted behavioural tests can reveal failures under controlled linguistic variation. \citet{gardner2020} developed contrast sets to examine whether meaningful changes in input produce appropriate changes in model predictions. Together, these approaches distinguish desirable consistency under irrelevant variation from sensitivity to information that should affect the outcome.

LLM-based evaluation introduces further instability. \citet{anghel2025} investigated positional bias and inconsistency in pairwise judgments, while \citet{zhu2026} reported differences in test–retest reliability, inter-model agreement and rating-scale utilisation during automated behavioural interview scoring. The latter remains a preprint and should therefore be interpreted accordingly. \citet{hua2025} additionally demonstrated that apparent prompt sensitivity may arise partly from evaluation procedures that fail to recognise semantically acceptable responses.

These findings are relevant to competency assessment because linguistic presentation may influence automated recommendations even when substantive evidence appears unchanged. However, differences in expression cannot automatically be dismissed as irrelevant: communication clarity may itself be an assessed competency. Stability evaluation must therefore distinguish meaning-preserving variation from changes that affect the evidence available to the evaluator. ESDSA operationalises this distinction through separately annotated semantic-equivalence groups, while recognising that the annotations themselves may be uncertain.

\subsection{Decision Validity, Uncertainty and Abstention}
Abstention has an established foundation in statistical classification. \citet{chow1970} formalised the relationship between recognition error and rejection, while \citet{elyaniv2010} examined the foundations of selective classification. \citet{geifman2017} subsequently investigated selective classification in deep neural networks, before introducing SelectiveNet, an architecture incorporating rejection directly into model learning \citep{geifman2019}. These contributions establish that withholding uncertain predictions is not inherently novel and must be evaluated in relation to both retained-decision reliability and coverage.

In employment assessment, the conditions justifying abstention are not necessarily captured by predictive confidence alone. \citet{castleman2026} investigated validity in LLM-based resume screening through controlled comparisons of candidates and qualifications, providing a precedent for examining whether automated selection responds appropriately to job-relevant information. Interview responses introduce additional complexity because partial accounts of reasoning or professional practice may not support either a favourable or unfavourable competency judgment.

ESDSA differs from confidence-based selective prediction by imposing explicit procedural requirements for actionability. Its criterion-level evidence classifications determine whether the frozen decision rules are satisfied, deterministic quotation matching checks textual support, and semantic-equivalence assessment governs the stability requirement. These mechanisms do not estimate a calibrated probability of correctness and cannot exclude confident but incorrect assessments, a broader concern examined by \citet{lakkaraju2017}. Nor does their integration establish superior decision quality. A more restrictive procedure may reduce reference disagreement simply by deferring additional responses, including those that previously agreed with the reference. Consequently, actionable coverage and the composition of deferrals must accompany any comparison of disagreement rates.

\subsection{Research Gap and Proposed Contribution}
Existing scholarship addresses several components of automated assessment, but their relationship within consequential competency recommendations remains an important methodological question. \citet{tripathi2026} identified methodological variation across research on LLM-based hiring, while \citet{raghavan2020} highlighted the need to scrutinise assessment validation claims. Selective prediction provides established rejection mechanisms \citep{chow1970,geifman2017}, behavioural testing examines sensitivity to linguistic changes \citep{ribeiro2020,gardner2020}, and research on explanation faithfulness questions whether plausible model-generated justifications accurately represent the basis of predictions \citep{turpin2023}. None of these approaches independently establishes when a particular interview recommendation satisfies both evidential and stability requirements.

ESDSA contributes an interview-specific decision-assurance protocol that integrates three previously established methodological ideas into a single operational procedure: criterion-level evidence classification, deterministic verification of supporting quotations, and semantic-equivalence-conditioned decision stability. Its distinctive contribution is not any individual mechanism, but the explicit specification of their combined conditions for actionability and the ability to examine their consequences against simpler evaluation approaches. Empirically, the study contributes a structured synthetic benchmark and comparative evaluation showing how these requirements affect actionable coverage, provisional reference disagreement and deferral. The results also identify a limitation of unanimity-based stability enforcement: a single unresolved linguistic variant can cause otherwise actionable, reference-consistent recommendations to be withheld. These findings characterise the procedural costs of decision assurance rather than demonstrating improved hiring accuracy.

This is a methodological contribution rather than a claim to have invented abstention, evidence-based assessment or linguistic robustness testing. The study evaluates a specified combination of existing ideas under controlled synthetic conditions, without independently validated hiring outcomes. Importantly, its empirical comparisons allow the integrated framework to perform less favourably than simpler approaches. The research gap is therefore addressed through an explicit examination of the conditions and costs of automated actionability, rather than an assumption that additional safeguards necessarily improve decisions.

\section{Research Questions and Hypotheses}
Automated interview recommendations raise two related questions: whether the available response contains sufficient competency evidence and whether the resulting assessment remains stable across substantively equivalent linguistic presentations. This study investigates these conditions through three research questions:

\noindent\textbf{RQ1 --- Evidence Sufficiency:} How do LLM-based interview evaluators distinguish positive, negative and insufficient competency evidence, and to what extent do their actionable recommendations agree with provisional rubric-based reference judgments?\par
\noindent\textbf{RQ2 --- Decision Stability:} Do automated recommendations differ across linguistic variants assessed as substantively equivalent, and how does applying a predefined stability requirement affect actionable decisions and deferrals?\par
\noindent\textbf{RQ3 --- Combined Assurance:} How does integrating evidence sufficiency, quotation traceability and semantic-equivalence stability affect actionable coverage, provisional reference disagreement and deferral compared with simpler evaluation approaches?\par
The investigation was motivated by three expectations:

\noindent\textbf{H1 --- Evidence Sufficiency:} Automated evaluators may issue actionable recommendations despite insufficient evidence under predefined competency criteria.\par
\noindent\textbf{H2 --- Decision Stability:} Linguistic variants assessed as substantively equivalent may produce different automated recommendations.\par
\noindent\textbf{H3 --- Combined Assurance:} Integrating evidence, traceability and stability requirements may reduce reference disagreement relative to evidence-aware evaluation, while increasing deferrals.\par
These expectations are examined descriptively rather than treated as preregistered confirmatory hypotheses. In particular, H3 was refined after the completed results were available and must therefore be regarded as a retrospective analytical expectation, not an independently tested prospective prediction.

Provisional decisions were derived from competency rubrics and AI-assisted annotations, without independent human ground-truth adjudication. An unsupported ADVANCE denotes an ADVANCE recommendation that disagrees with the provisional reference, rather than a verified assessment error. Accordingly, the analysis considers reference disagreement alongside actionable coverage and the composition of additional deferrals. The objective is to characterise the consequences of evidence-centred decision constraints, not to establish improvements in actual hiring accuracy or fairness.

\section{Methodology and Experimental Design}
\subsection{Research Design}
The study employed a controlled experimental design to examine evidence sufficiency and decision stability in LLM-based interview assessment. Synthetic responses to structured interview questions were evaluated across three occupational domains using four approaches: scoring-only evaluation, evidence-aware evaluation, Traceability v2 and integrated ESDSA v2. The analysis focused on whether recommendations satisfied predefined evidential and stability requirements, rather than whether they predicted actual employment suitability.

Responses were generated using GPT-5.4-mini and assessed against competency-specific rubrics. The evaluation distinguished three experimental recommendations: ADVANCE, DO\_NOT\_ADVANCE and DEFER. Their agreement with AI-assisted provisional references was examined alongside actionable coverage and deferral, without treating those references as independently validated ground truth.

\subsection{Assessment Domains and Benchmark Construction}
The benchmark covered Software Engineering (SWE), Customer Support (CSS) and Operations (OPS), each represented by four competencies. Every competency was operationalised through three assessment criteria, producing 12 rubrics and 36 criteria across the three domains.

The scenario registry contained 120 families, allocated as 72 DEVELOPMENT, 24 VALIDATION and 24 TEST families. However, only 12 DEVELOPMENT families were included in the initial evaluation, alongside all 24 families in each subsequent partition. The completed experiments therefore examined 60 scenario families and 540 synthetic interview responses.

Each evaluated family comprised three intended evidence conditions—positive, negative and insufficient—with three linguistic variants per condition: CANONICAL, CONVERSATIONAL and CONCISE. This produced nine responses and three semantic-equivalence groups per family. In TEST, eight families represented each occupational domain, with two families for each competency.

The three evidence conditions were designed to vary the information demonstrated in the response. Positive-condition responses described relevant competency evidence, negative-condition responses introduced materially deficient practice, and insufficient-condition responses withheld information required for either judgment. These conditions reflected generation intentions rather than independently verified measures of professional ability.

\subsection{Evidence Classification and Controlled Linguistic Variation}
Criterion-level annotations distinguished POSITIVE\_EVIDENCE, NEGATIVE\_EVIDENCE and INSUFFICIENT\_EVIDENCE. Positive evidence indicated that a response demonstrated the relevant assessment criterion. Negative evidence required an explicit demonstration of materially deficient practice, while insufficient evidence indicated that neither classification was adequately established. Missing evidence was therefore not automatically interpreted as deficient competence.

Within each intended evidence condition, the response generator produced three linguistic variants designed to preserve the same substantive competency information while varying expression, organisation and detail. Because the generation instructions could not guarantee equivalence, separate AI-assisted annotations classified each group as YES, NO or UNCERTAIN.

This design permitted two related comparisons. Differences across intended evidence conditions were used to examine how recommendations responded to changes in presented competency information. Comparisons within groups annotated as equivalent examined whether substantively similar responses received consistent decisions. Neither comparison independently established candidate competence or assessment validity.

\subsection{Evaluation Approaches}
Four evaluation approaches were compared:

The scoring-only baseline assigned each criterion a numerical score from 1 to 4. Under its original policy, a mean score of at least 3.0 produced ADVANCE, a mean of 1.5 or below produced DO\_NOT\_ADVANCE, and intermediate scores produced DEFER.

The evidence-aware baseline classified criterion-level information as positive, negative or insufficient evidence and applied its original evidence-aware decision rules. Its evaluation prompts and decision procedures were preserved across VALIDATION and TEST.

Traceability v2 applied the frozen ESDSA v2 quotation-verification and evidence decision procedures to the criterion-level assessments. Supporting excerpts were classified as VERIFIED, CONTEXT\_ONLY or UNVERIFIABLE, with NOT\_APPLICABLE recorded where quotation verification was not required. Verification used deterministic textual matching to establish whether supporting material occurred in the candidate's response. It did not establish that the excerpt substantively justified the evidence classification.

Integrated ESDSA v2 extended Traceability v2 through a semantic-equivalence stability gate. The gate conditioned actionability on consistent pre-stability decisions across linguistic variants annotated as substantively equivalent. Its policy was frozen before the final evaluation and was not modified in response to TEST results.

\subsection{Frozen ESDSA v2 Decision Policy}
ESDSA v2 applied three decision rules to the verified criterion-level evidence. A response received DO\_NOT\_ADVANCE when at least one criterion contained verified negative evidence. ADVANCE required verified positive evidence for all three criteria. Every remaining combination resulted in DEFER.

The policy deliberately distinguished insufficient evidence from demonstrated deficient practice. Its decision requirements were experimental rules for automated competency evaluation, not validated standards for employment selection. Textual verification also did not guarantee that the evaluator's interpretation of the quoted evidence was substantively correct.

The semantic stability procedure operated on groups of three linguistic variants. Where equivalence was annotated as YES, all three pre-stability recommendations had to be identical and actionable for those recommendations to remain actionable. Otherwise, the group was assigned DEFER. Where equivalence was NO or UNCERTAIN, individual pre-stability recommendations were retained without establishing semantic stability.

The framework therefore permitted deferral because criterion-level evidence remained unresolved or because a group of equivalent responses failed the specified stability requirement. Neither outcome established that the candidate lacked the assessed competency.

\begin{figure}[H]
    \centering
    \includegraphics[width=0.96\linewidth]{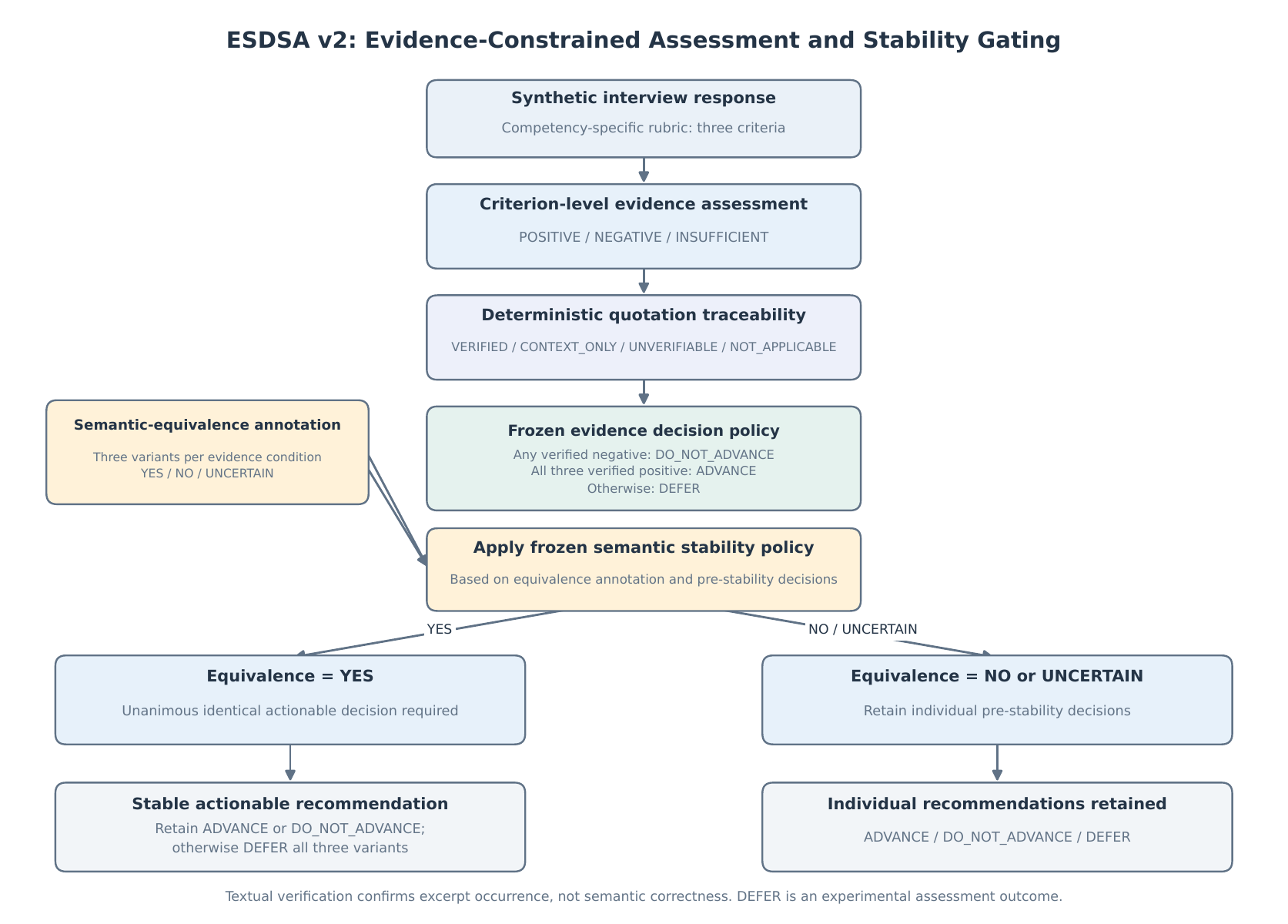}
    \caption{ESDSA v2 evidence assessment and semantic stability workflow. Candidate responses undergo criterion-level evidence assessment, deterministic quotation traceability and application of the frozen evidence decision policy. For groups assessed as semantically equivalent, actionable recommendations are retained only when all three linguistic variants produce identical actionable decisions; otherwise, all three are deferred. Groups classified as non-equivalent or uncertain retain their individual pre-stability decisions. Textual verification confirms excerpt occurrence, not the substantive correctness of evidence interpretation.}
    \label{fig:architecture}
\end{figure}

\subsection{Experimental Stages and Reference Preparation}
The DEVELOPMENT registry contained 72 scenario families, of which 12 (108 responses) were evaluated during the initial methodology-development stage. This preliminary subset was used to examine differences in actionable coverage, reference disagreement and decision consistency before finalising ESDSA v2; the available experimental records do not establish a formal sampling rationale for selecting these families. The subsequent refinement introduced deterministic quotation verification, explicit decision rules distinguishing verified negative, positive and insufficient evidence, and a semantic-equivalence stability gate requiring consistent actionable recommendations across equivalent linguistic variants. These procedures were frozen before VALIDATION and TEST, and the remaining 60 DEVELOPMENT families were not included in the reported experiments.

VALIDATION comprised 24 families, 216 responses, 648 criterion-level provisional evidence annotations and 72 semantic-equivalence assessments. The four approaches were compared using the frozen procedures without modifying the final v2 decision rules.

The TEST partition comprised a further 24 families and 216 responses. Response generation retained the original prompt and schema, with the returned model version recorded as gpt-5.4-mini-2026-03-17. The scoring-only and evidence-aware evaluators each produced 216 predictions. All 432 predictions were generated before reference comparison and subsequently fingerprinted and locked.

TEST reference preparation used AI-assisted criterion-level annotations and semantic-equivalence assessments. Of 216 response-level evidence annotations, 214 were accepted from the annotation procedure, while two required documented AI-assisted manual adjudication following repeated quotation-validation failures. These interventions affected six criterion judgments and did not constitute independent human adjudication.

Provisional reference decisions were prepared without reading the locked evaluator predictions. The final consolidated analysis subsequently applied the frozen traceability and stability procedures without additional model calls, policy modifications or threshold adjustments.

The TEST scenarios had nevertheless been authored after VALIDATION findings were examined. Although TEST predictions and reference outcomes had not yet been generated, scenario construction was therefore not entirely independent of the preceding analysis.

\subsection{Evaluation Measures and Analytical Approach}
The primary outcome was disagreement with AI-assisted pre-stability provisional reference decisions among actionable recommendations. Actionable coverage represented the proportion of responses receiving ADVANCE or DO\_NOT\_ADVANCE rather than DEFER.

Disagreement was calculated as the number of actionable recommendations differing from their corresponding provisional references divided by the total number of actionable recommendations. Because each approach retained a different subset of responses, conditional disagreement was interpreted jointly with coverage.

Unsupported ADVANCE was defined as an ADVANCE recommendation inconsistent with the provisional reference. It did not represent an independently established hiring error.

The stability-gate analysis distinguished additional deferrals according to whether the original actionable recommendation agreed or disagreed with its pre-stability reference. Secondary comparisons used post-stability provisional references, but these were not treated as independent validation because they incorporated the stability mechanism under evaluation.

The analysis was primarily descriptive. Responses were clustered within scenario families, with shared questions, evidence conditions and linguistic variants. Consequently, individual responses could not be assumed statistically independent, and differences between approaches were not interpreted as demonstrating statistically established superiority.

\subsection{Scope and Reproducibility}
The experiments used synthetic responses rather than actual applicants and did not evaluate employment outcomes, protected-group fairness or operational recruitment decisions. The provisional references were predominantly AI-assisted, with the same model family used for response generation and most reference annotations. These dependencies limited the interpretation of agreement as a measure of assessment correctness.

Reproducibility was supported through predefined competency rubrics, preserved baseline evaluation procedures, frozen ESDSA v2 policies, recorded model information, prediction-locking records and documented annotation interventions. The final TEST analysis was performed using the completed outputs without modifying the frozen decision procedures.

The study therefore evaluated the behaviour of specified evidence and stability requirements under controlled conditions. Broader limitations concerning reference validity, statistical dependence and real-world generalisation are examined in Section 7.

\section{Experimental Results}
\subsection{Development-Stage Findings}
The initial DEVELOPMENT evaluation examined 108 synthetic responses from 12 scenario families, including 36 groups of linguistic variants. These preliminary experiments informed refinement of the evidence and stability procedures rather than providing an independent assessment of generalisation.

Scoring-only evaluation produced 79 actionable recommendations (73.1\% coverage), compared with 105 (97.2\%) under the original evidence-aware approach. Disagreement with provisional references among actionable decisions was 10.1\% and 14.3\%, respectively. Higher coverage therefore coincided with greater conditional disagreement.

Decision instability also appeared across linguistic variants. Among 27 groups provisionally assessed as equivalent and assigned consistent reference decisions, scoring-only evaluation produced identical recommendations in 23 groups, compared with 25 under evidence-aware evaluation. In GROUP-017, concerning duplicate orders potentially arising from retry requests, the evidence-aware evaluator advanced one linguistic variant but rejected two others despite their intended substantive equivalence. This illustrated inconsistent interpretation of the candidate's investigation and proposed corrective action, subject to the limitations of the provisional equivalence annotations.

The refined integrated ESDSA v2 procedure subsequently produced 100 actionable recommendations (92.6\% coverage), including 13 disagreements (13.0\%), ten unsupported ADVANCE recommendations and eight deferrals. These developmental findings motivated the subsequent comparison of frozen procedures in VALIDATION and TEST.

\subsection{Validation Results}
VALIDATION comprised 24 scenario families and 216 responses, with 648 criterion-level provisional evidence judgments and 72 semantic-equivalence assessments. Table 1 summarises the four approaches against AI-assisted pre-stability provisional references.

\begin{table}[H]
\centering
\caption{VALIDATION comparison of the four approaches (216 synthetic responses). Disagreement is calculated among actionable recommendations against AI-assisted pre-stability provisional references. Unsupported ADVANCE denotes an ADVANCE recommendation inconsistent with the provisional reference, not an independently verified assessment error.}
\label{tab:validation}
\resizebox{\textwidth}{!}{\begin{tabular}{lcccc}
\toprule
Approach & Actionable coverage & Disagreement among actionable & Unsupported ADVANCE & Deferrals \\
\midrule
Scoring-only & 66.7\% & 6.2\% (9/144) & 6 & 72 \\
Evidence-aware & 91.7\% & 13.1\% (26/198) & 14 & 18 \\
Traceability v2 & 88.9\% & 13.5\% (26/192) & 14 & 24 \\
Integrated ESDSA v2 & 76.9\% & 9.0\% (15/166) & 4 & 50 \\
\bottomrule
\end{tabular}}
\end{table}

Evidence-aware evaluation achieved the highest actionable coverage (91.7\%), compared with 66.7\% for scoring-only assessment. However, its 26 disagreements among 198 actionable decisions (13.1\%) exceeded the nine disagreements among 144 scoring-only decisions (6.2\%). Traceability v2 retained 192 actionable decisions, with 26 disagreements (13.5\%), indicating that quotation verification alone did not reduce reference disagreement in this partition.

Integrated ESDSA v2 produced 166 actionable decisions (76.9\% coverage), with 15 disagreements (9.0\%) and four unsupported ADVANCE recommendations. Relative to evidence-aware evaluation, it reduced reference disagreement but increased deferrals from 18 to 50. Scoring-only evaluation nevertheless retained a lower conditional disagreement rate.

The stability gate introduced 26 additional deferrals relative to Traceability v2. Eleven previously disagreed with provisional references, whereas 15 had agreed. Of the 72 semantic-equivalence groups, 50 were annotated YES, 21 NO and one UNCERTAIN. These classifications determined which groups were subject to the stability requirement.

\subsection{Held-Out TEST Results}
TEST comprised 24 scenario families and 216 responses, representing all 12 competencies across the three occupational domains. The primary provisional references comprised 128 ADVANCE, 66 DO\_NOT\_ADVANCE and 22 DEFER decisions. All 432 baseline predictions had been locked before reference comparison.

\begin{table}[H]
\centering
\caption{Held-out TEST comparison of four automated interview evaluation approaches (216 synthetic responses). The primary outcome is disagreement among actionable recommendations against AI-assisted pre-stability provisional references. Coverage and disagreement must be interpreted jointly because methods retain different subsets of responses.}
\label{tab:test}
\resizebox{\textwidth}{!}{\begin{tabular}{lcccc}
\toprule
Approach & Actionable coverage & Disagreement among actionable & Unsupported ADVANCE & Deferrals \\
\midrule
Scoring-only & 71.30\% & 1.95\% (3/154) & 3 & 62 \\
Evidence-aware & 92.59\% & 4.50\% (9/200) & 6 & 16 \\
Traceability v2 & 90.28\% & 4.10\% (8/195) & 5 & 21 \\
Integrated ESDSA v2 & 85.19\% & 3.80\% (7/184) & 4 & 32 \\
\bottomrule
\end{tabular}}
\end{table}

Scoring-only evaluation produced 154 actionable decisions (71.30\% coverage), with three disagreements (1.95\%). Evidence-aware evaluation achieved higher coverage (92.59\%), but nine of its 200 actionable recommendations disagreed with provisional references (4.50\%). Traceability v2 retained 195 actionable decisions, with eight disagreements (4.10\%), representing five additional deferrals and one fewer disagreement relative to evidence-aware evaluation.

Integrated ESDSA v2 produced 184 actionable recommendations (85.19\% coverage), with seven disagreements (3.80\%) and four unsupported ADVANCE recommendations. Compared with evidence-aware evaluation, integration reduced disagreement from nine to seven while increasing deferrals from 16 to 32.

Disaggregated TEST analysis showed that reference disagreements were concentrated in responses generated under the intended INSUFFICIENT evidence condition. Evidence-aware evaluation produced seven disagreements among 58 actionable recommendations (12.1\%) in this condition, compared with five among 51 (9.8\%) for integrated ESDSA. Both approaches recorded two disagreements among 70 actionable recommendations (2.9\%) under the NEGATIVE condition and none under the SUFFICIENT condition. All six evidence-aware unsupported ADVANCE recommendations occurred in the intended INSUFFICIENT condition, compared with four under integrated ESDSA. By occupational domain, Customer Support accounted for five of nine evidence-aware disagreements and three of seven integrated disagreements. At competency level, CSS01 disagreements decreased from two to zero, accompanied by a reduction in actionable recommendations from 17 to 12. These descriptive differences identify concentrations of provisional reference disagreement but do not establish independently verified evidence insufficiency, since the generation conditions and reference annotations were not externally validated.

The integrated approach therefore occupied an intermediate position between the scoring-only and evidence-aware baselines. It did not achieve the lowest conditional disagreement rate, nor did it retain the highest proportion of actionable decisions. Figure 2 illustrates these coverage–disagreement relationships across VALIDATION and TEST.

\begin{figure}[H]
\centering
\includegraphics[width=0.96\linewidth]{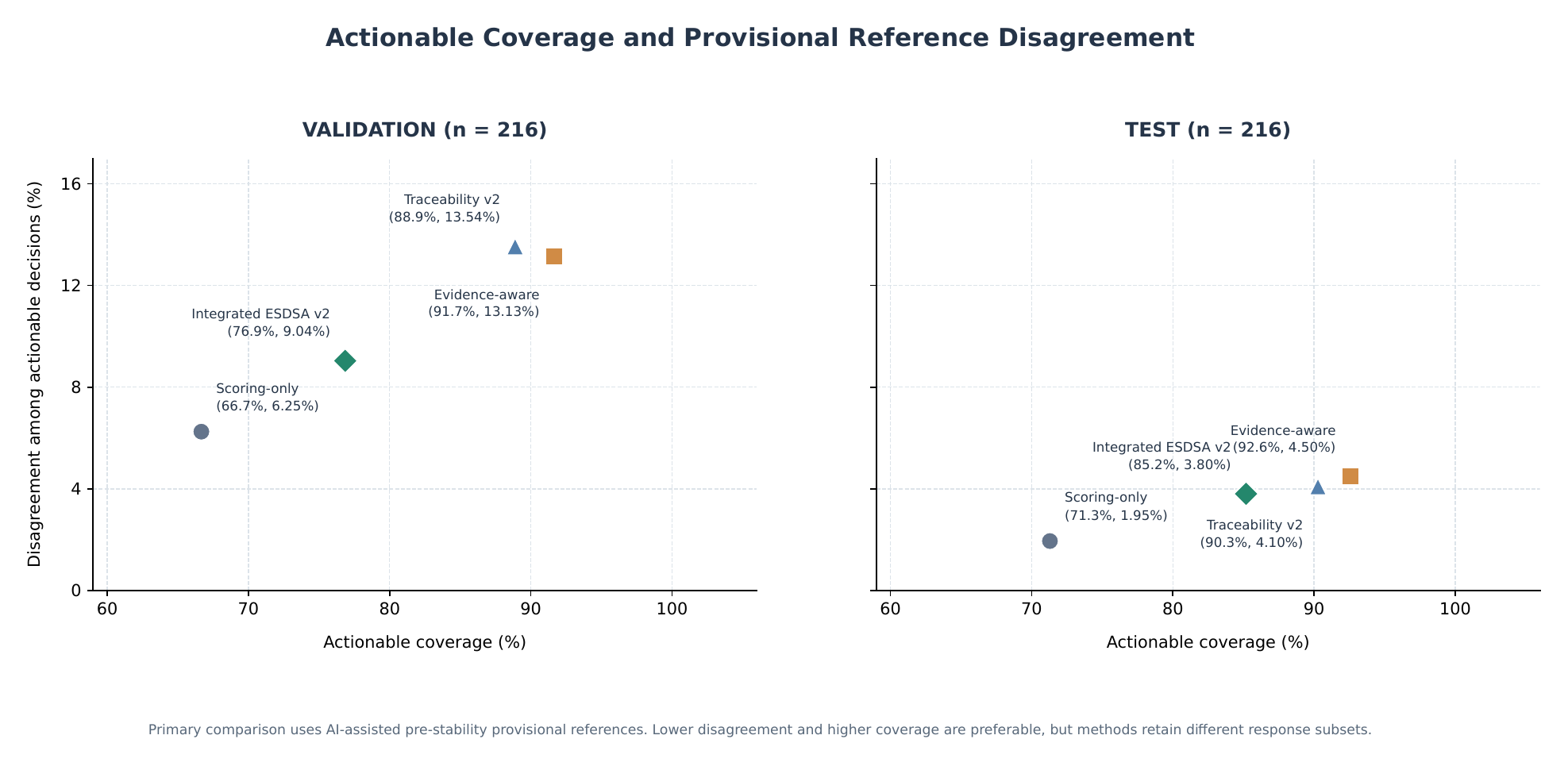}
\caption{Actionable coverage and disagreement with AI-assisted pre-stability provisional reference judgments across VALIDATION and held-out TEST. Each point represents one evaluation approach. Actionable coverage is the proportion of responses receiving ADVANCE or DO\_NOT\_ADVANCE; disagreement is calculated among actionable recommendations. Because the approaches retain different subsets of responses, the plotted disagreement rates should not be interpreted as independently validated classification accuracy.}
\label{fig:figure2}
\end{figure}

\subsection{Stability-Gate Effects and Traceability Outcomes}
The contribution of the stability gate was examined by comparing Traceability v2 decisions with those retained after semantic-equivalence assessment. Table 3 separates additional deferrals according to their agreement with pre-stability provisional references.

\begin{table}[H]
\centering
\caption{Effects of the frozen semantic-equivalence stability gate in VALIDATION and TEST. Agreement and disagreement are measured against pre-stability AI-assisted provisional references. Previously reference-consistent decisions are not necessarily independently justified or correct.}
\label{tab:stability}
\resizebox{\textwidth}{!}{\begin{tabular}{lrr}
\toprule
Measure & VALIDATION & TEST \\
\midrule
Actionable decisions before stability gate & 192 & 195 \\
Actionable decisions after stability gate & 166 & 184 \\
Additional deferrals & 26 & 11 \\
Deferred decisions previously disagreeing with references & 11 & 1 \\
Deferred decisions previously agreeing with references & 15 & 10 \\
Total deferrals after integration & 50 & 32 \\
\bottomrule
\end{tabular}}
\end{table}

In VALIDATION, the gate deferred 26 previously actionable decisions, including 11 that had disagreed with provisional references and 15 that had agreed. In TEST, it deferred 11 additional decisions, of which only one had disagreed with its reference. The other ten had previously been reference-consistent. Consequently, the reduction in actionable disagreement attributable to the stability gate was considerably smaller in TEST. Regarding RQ2 and H2, the DEVELOPMENT findings identified differences in recommendations across some groups assessed as semantically equivalent. The subsequent VALIDATION and TEST analyses quantified the decisions withheld by the frozen stability gate rather than reporting a separate frequency of inconsistent groups. The additional deferral counts should therefore be interpreted as the consequences of applying the stability requirement, not as direct estimates of linguistic instability.

A case-level examination identified a common pattern underlying all ten additional reference-consistent TEST deferrals. These decisions originated from five semantic-equivalence groups, each containing two variants assigned ADVANCE and a third assigned DEFER before application of the stability gate. All three variants in each group had provisional reference judgments of ADVANCE. The frozen unanimity requirement consequently withheld both previously actionable recommendations in every group. Three groups belonged to the intended SUFFICIENT condition and two to INSUFFICIENT, spanning Customer Support, Operations and Software Engineering. This pattern demonstrates how a single unresolved variant can propagate deferral to otherwise actionable responses under the specified policy. However, the recorded decisions do not establish whether the initial variation resulted from linguistic sensitivity, differences in evidence interpretation or annotation uncertainty.

TEST semantic-equivalence annotations classified 46 groups as YES, 25 as NO and one as UNCERTAIN. Only groups classified YES were subject to the frozen stability requirement, making the resulting deferrals dependent on the provisional equivalence annotations.

The TEST traceability audit recorded 608 VERIFIED, 34 NOT\_APPLICABLE, five UNVERIFIABLE and one CONTEXT\_ONLY classifications across 648 criterion-level assessments. Relatively few assessments therefore failed the textual verification requirements. However, VERIFIED established excerpt occurrence rather than the substantive correctness of evidence interpretation.

These findings indicate that stability-based deferral should not be equated with error identification. The gate enforced a specified consistency requirement, including where the recommendations it withheld had previously agreed with provisional references.

\subsection{Secondary Reference Comparison and Summary}
A secondary TEST analysis used post-stability provisional references, comprising 124 ADVANCE, 64 DO\_NOT\_ADVANCE and 28 DEFER decisions. Against these references, scoring-only evaluation produced four disagreements among 154 actionable recommendations, evidence-aware evaluation 15 among 200, Traceability v2 14 among 195 and integrated ESDSA v2 13 among 184.

These outcomes were not independently validated stability measures because the post-stability references incorporated the same mechanism examined by integrated ESDSA. The pre-stability comparisons therefore remained primary.

Regarding RQ1 and H1, the experiment operationalised positive, negative and insufficient evidence through criterion-level classifications, but the reported aggregate results do not separately quantify actionable recommendations issued despite insufficient criterion-level evidence. The unsupported ADVANCE measure instead captures disagreement with provisional reference judgments. Consequently, the findings characterise evidence-aware decision behaviour and reference agreement but do not constitute a direct empirical confirmation of H1.

Across VALIDATION and TEST, evidence-aware evaluation achieved the highest actionable coverage, while scoring-only evaluation produced the lowest conditional reference disagreement. Integrated ESDSA reduced disagreement relative to evidence-aware evaluation at the expense of additional deferrals, without demonstrating uniform superiority.

The central empirical finding is that evidence constraints and semantic stability requirements alter which recommendations remain actionable, but do not necessarily improve agreement with provisional references. In TEST, most additional stability-based deferrals concerned previously reference-consistent decisions. This limits claims that the stability mechanism selectively identifies erroneous assessments while establishing its measurable effects on coverage and abstention.

\section{Discussion}
\subsection{Evidence Sufficiency and the Limits of Automated Assessment}
The findings distinguish an evaluator's ability to issue an actionable recommendation from the evidential justification of that recommendation. The four approaches produced different coverage and disagreement profiles despite assessing the same synthetic responses. These differences reflect their respective decision policies, not independently established differences in the ability to identify suitable candidates.

This distinction is relevant to construct validity. As \citet{raghavan2020} demonstrate, algorithmic hiring assessments require scrutiny of the relationship between claimed capabilities and supporting validation evidence. ESDSA addresses a narrower question: whether the information presented in an interview response satisfies predefined conditions for actionability. An incomplete answer may leave a competency unresolved without demonstrating deficient professional practice.

Nevertheless, criterion-level evidence classification does not resolve the broader validity problem. Evidence-aware evaluation achieved the highest coverage in both partitions but also generated more provisional reference disagreements than the other approaches. Conversely, scoring-only evaluation produced fewer disagreements while deferring substantially more responses. Neither result establishes which approach would better predict workplace performance. The contribution of evidence sufficiency is therefore procedural: it makes unresolved evidence explicit without claiming that the underlying competency has been independently measured.

\subsection{Abstention and the Coverage-Disagreement Trade-Off}
Selective prediction provides an established basis for understanding the relationship between actionable coverage and error \citep{geifman2017,geifman2019}. ESDSA extends this discussion to procedural conditions for issuing competency recommendations, although its deferral mechanism does not estimate calibrated predictive risk.

The TEST findings illustrate the importance of evaluating coverage and disagreement jointly. Scoring-only assessment achieved the lowest conditional disagreement (1.95\%) but retained only 71.30\% of responses as actionable. Evidence-aware evaluation achieved 92.59\% coverage with 4.50\% disagreement, while integrated ESDSA occupied an intermediate position at 85.19\% coverage and 3.80\% disagreement.

These rates cannot establish that greater selectivity produces better assessments. Each approach retained a different subset of responses, changing the cases included in its disagreement calculation. Lower disagreement may reflect the removal of difficult assessments rather than improved interpretation of the remaining evidence.

The relevant contribution of ESDSA is therefore not abstention itself, but the explicit specification of the conditions triggering it. Whether these conditions offer a better trade-off than confidence-based rejection or alternative evidence policies remains unresolved.

\subsection{Quotation Traceability and Evidential Justification}
The distinction between traceability and substantive justification is central to interpreting the results. Research on explanation faithfulness demonstrates that plausible model-generated explanations need not accurately represent the factors influencing a prediction \citep{turpin2023}. ESDSA does not establish explanation faithfulness in this broader sense. Its deterministic procedure verifies whether supporting excerpts appear in the candidate's response.

The modest effect of Traceability v2 in TEST is consistent with this limited function. Only six of the 648 criterion-level assessments received UNVERIFIABLE or CONTEXT\_ONLY classifications, and traceability reduced reference disagreements from nine to eight relative to evidence-aware evaluation. Most excerpts satisfied the textual matching requirement, but this does not establish that they were correctly interpreted.

An authentic quotation can still be irrelevant, ambiguous or insufficient to support the assigned classification. Traceability consequently strengthens the inspectability of an assessment without independently validating the inference drawn from it. Future assurance procedures would need to examine not only whether supporting material exists, but whether its interpretation is substantively defensible.

\subsection{Semantic Stability and the Cost of Additional Deferral}
Decision stability offers a further procedural condition for actionability. Where linguistic variants preserve substantive competency evidence, divergent recommendations may indicate sensitivity to presentation rather than relevant differences in professional reasoning. However, stable decisions can still be incorrect, and apparent instability may arise from errors in semantic-equivalence annotation.

The frozen stability gate exposed an important limitation of consistency-based deferral. In VALIDATION, 11 of 26 additional deferrals concerned recommendations that had disagreed with provisional references. In TEST, only one of 11 additional deferrals concerned such a disagreement; the remaining ten had previously been reference consistent.

The TEST findings therefore provide limited support for interpreting the stability gate as a mechanism that selectively identifies reference-disagreeing recommendations. Its clearer demonstrated function is enforcing an additional consensus requirement, even where individual recommendations previously agreed with the reference. These deferrals cannot be labelled unnecessary without independent evidence of correctness, but neither can they be assumed beneficial.

Figure 3 illustrates this contrast between the two evaluation partitions.

\begin{figure}[H]
\centering
\includegraphics[width=0.96\linewidth]{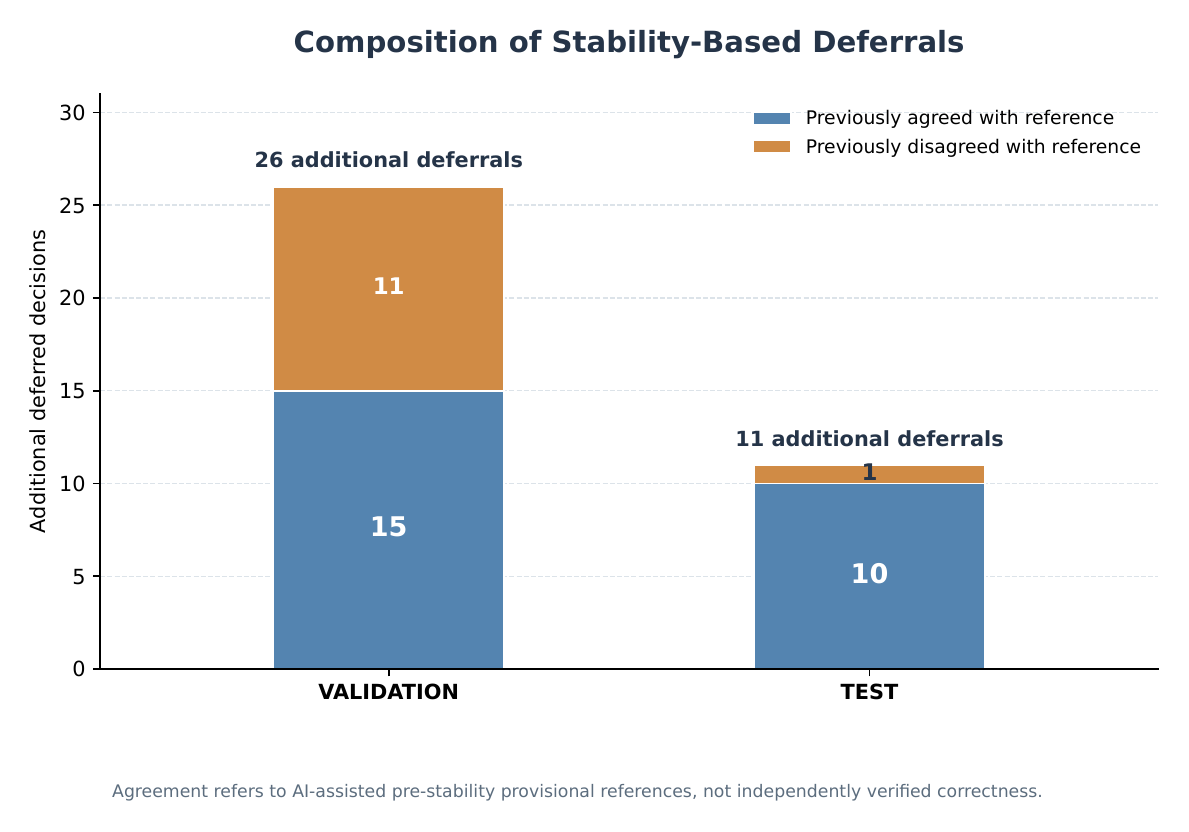}
\caption{Composition of additional stability-based deferrals in VALIDATION and held-out TEST. The stacked bars distinguish previously actionable decisions that agreed or disagreed with AI-assisted pre-stability provisional references before application of the frozen semantic-equivalence stability gate. Reference agreement does not establish decision correctness, and the figure does not imply that all reference-consistent deferrals were unnecessary.}
    \label{fig:figure3}
\end{figure}

The mechanism also depends on the reliability of the equivalence annotations. In TEST, 46 of 72 groups were classified as equivalent, while 25 were non-equivalent and one uncertain. Only equivalent groups underwent the stability gate. Incorrect equivalence judgments could therefore either trigger inappropriate deferral or prevent a genuine inconsistency from being examined. Stability should consequently be understood as a defined procedural property rather than independent evidence of assessment validity.

\subsection{Provisional References and Evaluator Dependence}
The interpretation of all comparative findings depends on the quality of the provisional reference judgments. These were predominantly AI-assisted and were not independently adjudicated by occupational experts. The same model family also generated the synthetic responses and produced most reference annotations, introducing potential shared interpretive tendencies.

Research on LLM evaluators identifies biases and sensitivities that complicate their use as independent assessment authorities \citep{liu2023,zheng2023}. Although predefined rubrics, frozen policies and locked TEST predictions reduced opportunities for retrospective adjustment, they could not eliminate reference error or evaluator dependence.

TEST scenario construction also followed examination of VALIDATION findings. The final predictions and references were protected from direct post hoc policy adjustment, but the scenario design was not fully independent of earlier observations.

Disagreement with these references is therefore best interpreted as consistency with a specified annotation procedure. Neither reference agreement nor disagreement establishes the correctness of an employment assessment.

\subsection{Implications for AI Assurance and Employment Decision Governance}
The findings support distinguishing the production of an automated assessment from the institutional decision to act upon it. Research on algorithmic hiring and machine-learning harms highlights the importance of examining validation claims, deployment assumptions and organisational processes rather than model outputs alone \citep{selbst2019,raghavan2020,suresh2021}.

ESDSA provides one operational approach to making evidence classifications, supporting excerpts and consistency requirements inspectable, without resolving the broader questions of fairness and measurement discussed by \citet{barocas2023}. Its empirical value lies in exposing the consequences of those requirements, including decisions withheld despite agreement with provisional references.

However, DEFER is not equivalent to effective human oversight. In practice, deferral would require a defined pathway for obtaining additional evidence, assigning responsibility for reassessment and allowing questionable judgments to be challenged. Without such arrangements, automated abstention may merely transfer unresolved uncertainty elsewhere in recruitment.

The frozen ESDSA policy should not be treated as a universally appropriate hiring standard, particularly given the range of language-model risks requiring contextual evaluation \citep{weidinger2022}. Competency requirements, the consequences of incorrect decisions and the available opportunities for further assessment may justify different rules. What the present study establishes is a reproducible way to examine the procedural conditions governing automated actionability and the costs those conditions impose. Whether they improve recruitment decisions remains a question for independently validated operational research.

\section{Limitations and Threats to Validity}
\subsection{Construct and Reference Validity}
The experimental rubrics assessed evidence presented in interview responses rather than independently validated professional competence. A response could satisfy a criterion without demonstrating workplace ability, while an incomplete answer might reflect limited opportunity or interpretation rather than deficient practice.

Provisional references were predominantly AI-assisted, without independent occupational-expert adjudication. The same model family generated the synthetic responses and most reference annotations, potentially introducing shared interpretive biases. Two TEST responses required documented AI-assisted manual adjudication, affecting six criterion-level judgments. Accordingly, reference agreement cannot establish decision correctness or recruitment validity.

Inter-annotator reliability was not independently measured for either criterion-level evidence classifications or semantic-equivalence assessments. Although the annotation procedure included quotation-validation checks and documented interventions, these checks did not establish agreement between independent assessors. The reliability of the provisional annotations therefore remains an unresolved measurement limitation, particularly because errors in evidence classification or equivalence judgments could directly influence the reported disagreements and stability-based deferrals.

\subsection{Internal Validity and Experimental Independence}
Although the ESDSA v2 procedures were frozen and TEST predictions locked before reference comparison, TEST scenario questions were developed after VALIDATION results had been examined. The final evaluation was therefore protected against direct policy adjustment using TEST outcomes but was not fully independent of earlier experimental observations.

The study also evaluated one frozen decision policy rather than systematically comparing alternative evidence thresholds, quotation-verification rules or stability requirements. Its findings consequently describe the implemented configuration, not an optimal assessment policy.

\subsection{Statistical and Measurement Limitations}
The 216 TEST responses were organised within 24 scenario families, with shared questions, evidence conditions and linguistic variants. These observations were not statistically independent, and their number should not be interpreted as 216 independent experimental units.

Disagreement was calculated conditionally among actionable recommendations, with each approach retaining different subsets of responses. Consequently, the reported rates do not provide directly comparable estimates of general classification accuracy. The descriptive results do not establish statistically significant superiority or independently verified reductions in assessment error.

\subsection{Semantic Equivalence and Traceability}
The generation instructions did not guarantee that linguistic variants preserved identical competency evidence. AI-assisted equivalence annotations could therefore incorrectly classify meaningful differences as equivalent or overlook genuine inconsistencies.

Deterministic quotation matching verified textual occurrence but could not establish whether excerpts were relevant, correctly interpreted or sufficient to justify the resulting assessment. Both mechanisms provide procedural checks rather than independent evidence of substantive validity.

\subsection{External Validity and Operational Application}
All responses were synthetic and limited to Software Engineering, Customer Support and Operations. The experiments involved no actual applicants, employers or workplace performance outcomes. They therefore cannot establish predictive validity, protected-group fairness, candidate experience or compliance with employment requirements.

The study also did not implement the processes that would follow DEFER, such as human reassessment, additional questioning or candidate challenge. Its administrative and distributional consequences remain unknown.

ESDSA should therefore be understood as a controlled evaluation methodology rather than a validated recruitment system. Establishing its operational value requires independent reference assessment, real-world validation and examination of the decision processes following automated deferral.

\section{Conclusion}
Automated interview assessment raises a question beyond scoring consistency: whether the evidence presented in a candidate's response is sufficient to justify an actionable recommendation. This study introduced the Evidence Sufficiency and Decision Stability Assessment (ESDSA), a reproducible framework combining criterion-level evidence classification, deterministic quotation traceability and stability requirements across linguistic variants assessed as substantively equivalent.

The framework was evaluated using 540 synthetic interview responses across three occupational domains. In held-out TEST, integrated ESDSA achieved 85.19\% actionable coverage and 3.80\% disagreement among actionable recommendations against AI-assisted provisional references. Although it reduced disagreement relative to evidence-aware evaluation, the scoring-only baseline achieved a lower conditional disagreement rate at substantially lower coverage. Moreover, ten of the eleven additional TEST deferrals introduced by the stability gate concerned recommendations that had previously agreed with provisional references.

These findings demonstrate that imposing evidential and consistency requirements changes the distribution of actionable decisions without necessarily improving the coverage–disagreement trade-off. ESDSA's contribution is therefore methodological: it provides explicit, auditable conditions for distinguishing insufficient evidence from demonstrated deficient practice and identifying recommendations that fail specified stability requirements.

The results remain limited by synthetic responses, AI-assisted reference judgments and the absence of independently validated employment outcomes. Future research must establish whether these procedural safeguards improve substantive assessment validity and examine the consequences of deferral within operational recruitment processes.

\section*{Data and Code Availability}
The synthetic interview benchmark, competency rubrics, evaluation prompts, frozen ESDSA v2 policies, AI-assisted reference annotations and analysis code are publicly available at https://github.com/khalidaals/ESDSA. These materials support reproducibility of the experimental procedures and reported findings.

\clearpage
\bibliographystyle{unsrtnat}
\bibliography{references}
\end{document}